\documentclass[universe,article,accept,moreauthors]{Definitions/mdpi}
\UseRawInputEncoding
\usepackage{amsmath}
\usepackage{amsfonts}
\usepackage{amssymb,bm}
\usepackage{siunitx}
\usepackage{color}
\usepackage{braket}

\firstpage{1}
\pubvolume{7}
\issuenum{9}
\articlenumber{343}
\pubyear{2021}
\copyrightyear{2020}
\externaleditor{Academic Editor: Gerald B. Cleaver} 
\history{Received: 16 August 2021; Accepted: 9 September; Published: 12 September 2021}

\Title{Dark Matter Axions, Non-Newtonian Gravity and Constraints on them from
Recent Measurement of the Casimir Force in the Micrometer Separation Range}

\Author{
Galina L. Klimchitskaya $^{1,2}$,
Vladimir M. Mostepanenko $^{1,2,3}$
}

\address{%
$^{1}$  \quad Central Astronomical Observatory at Pulkovo of the Russian Academy of Sciences, Saint Petersburg, 196140, Russia\\
$^{2}$ \quad Institute of Physics, Nanotechnology and
Telecommunications, Peter the Great Saint Petersburg
Polytechnic University, Saint Petersburg, 195251,  Russia\\
$^{3}$ \quad Kazan Federal University, Kazan, 420008, Russia
}

\corres{Correspondence: vmostepa@gmail.com}

\abstract{We consider axionlike particles, as the most probable constituents
of dark matter, the Yukawa-type corrections to Newton's gravitational law
and constraints on their parameters following from astrophysics and different
laboratory experiments. After a brief discussion of the results by
Prof.\ Yu.\ N.\ Gnedin in this field, we are coming to the recent experiment
on measuring the differential Casimir force between Au-coated surfaces of a
sphere and the top and bottom of rectangular trenches. In this experiment,
the Casimir force was measured over an unusually wide separation region from
0.2 to $8~\mu$m and compared with the exact theory based on first principles
of quantum electrodynamics at nonzero temperature. We use the measure
of agreement between experiment and theory for obtaining constraints on the
coupling constant of axionlike particles to nucleons and on the interaction
strength of Yukawa-type interaction. The obtained constraints on the
axion-to-nucleon coupling constant and on the strength of Yukawa interaction
are stronger by the factors of 4 and 24, respectively, than those found
previously from gravitational experiments and measurements of the Casimir
force, but weaker than the constraints following from the
differential measurement where the Casimir
force was nullified. Some other already performed and planned experiments
directed to the search of axions and non-Newtonian gravity are discussed
and their prospects are evaluated.
}


\begin{document}
\section{Introduction}\label{Intro}

The problem of dark matter has a long history \cite{1}.
As was found by J.~Oort in 1932 when studying stellar motion in the neighborhood
of a galaxy, the galaxy mass must be well over than that of its visible
constituents \cite{2}. A year later, F.~Zwicky \cite{3} applied the virial theorem
to the Coma cluster of galaxies in order to determine its mass. The obtained mass
value turned out to be much larger than that found from the number of observed
galaxies belonging to the Coma cluster multiplied by their mean mass.
An excess mass, which reveals itself only gravitationally, received the name
{\it dark matter}.

According to current concepts, the dark matter contributes approximately 27\%
to the energy of the Universe although its physical nature remains unknown.
There are many approaches to this problem based on the role of some hypothetical
particles, such as axions, arions, massive neutrinos, weakly interacting massive
particles (WIMP), barionic dark matter, modified gravity etc.
(see \cite{1,4,5,6,7,8} for a review).

The model of dark matter which finds a support from astrophysics and cosmology
is referred to as {\it cold dark matter}.
According to this model, dark matter consists of light hypothetical particles which
are produced in the early Universe and become nonrelativistic already at the first
stages of its evolution. The most popular particle of this kind is an axion, i.e.,
a pseudoscalar Nambu-Goldstone boson introduced to solve the problem of strong
CP violation in Quantum Chromodynamics (QCD) \cite{9,10,11}.
It has been known that the gauge invariant QCD vacuum depends on an angle $\theta$,
and this dependence violates the CP invariance of QCD. However, experiment says
that strong interactions are CP invariant and the electric dipole moment of a neutron
is equal to zero up to a high degree of accuracy. An introduction of the Peccei-Quinn
symmetry and axions, which are connected with its violation, helps to solve this
problem.

Axions and other axionlike particles, which arise in many extensions to the Standard
Model, can interact both with photons and with fermions (electrons and nucleons).
These interactions are used for an axion search and for constraining axion masses and
coupling constants from observations of numerous astrophysical and cosmological
processes, as well as from various laboratory experiments (see reviews
\cite{1,4,5,6,7,8,12,13,14,15,16,17,18,19,20} of already obtained bounds on the axion
mass and coupling constants to photons, electrons, and nucleons).

Prof.~Yuri~N.~Gnedin obtained many important results investigating the interaction
of dark matter axions with photons in astrophysics and cosmology. He proposed
\cite{21} to employ
the polarimetric methods for a search for axions and arions (i.e., the axions of
zero mass \cite{22}) in the emission from pulsars, X-ray binaries with low-mass
components, and magnetic white dwarfs. For this purpose, it was suggested to use
the conversion process of photons into axions in the magnetic field of compact stars
and in the interstellar and intergalactic space (i.e., the Primakoff effect).
Next, Prof.\ Gnedin demonstrated an appearance of the striking feature in
the polarized
light of quasistellar objects due to the resonance magnetic conversion of photons into
    massless axions \cite{23}.

Using these results, Prof.~Yu.~N.~Gnedin organized the axion search by the 6-m
telescope at the Special Astronomical Observatory in Russia. Both the Primakoff effect
and the inverse process of an axion decay into two photons were searched in the
integalactic light of clusters of galaxies and in the brightness of night sky due to
axions in the halo of our Galaxy \cite{24}. Although no evidences of axions were
found, it was possible to find the upper limit on the photon-to-axion coupling
constant from the polarimetric observations of magnetic chemically peculiar stars of
spectral type A possessing strong hydrogen Balmer absorption lines \cite{24}.
The above results, as well as the ground-based cavity experiments searching for
galactic axions, searches of an axion decay in the galactic and extragalactic light,
for the solar and stellar axions, and the obtained limits on the coupling constant
of axions to photons, were discussed in the review \cite{25}.

In the further research of dark matter axions, Yu.~N.~Gnedin and his collaborators
analyzed \cite{26} the intermediate results of PVLAS experiment interpreted
\cite{27} as arising due to a conversion of photons into axions with a coupling
constant to photons of the order of $4\times 10^{-6}~\mbox{GeV}^{-1}$.
By considering the astrophysical and cosmological constraints, they have shown
\cite{26} that this result is in contradiction with the data on stellar evolution
that exclude the standard model of QCD axions.

Using the cosmic orientation of the electric field vectors of polarized radiation
from distant quasars,  Yu.~N.~Gnedin and his collaborators placed rather strong
limit on the coupling constant of axions to electric field \cite{28}.
Numerous results related to the processes of axion decay into two photons, the
transformation of photons into axions in the magnetic fields of stars and of
interstellar or intergalactic media, and the transformation of axions generated
in the cores of stars into X-ray photons were discussed in the review \cite{29}.

It has been known that the coupling constant of axionlike particles to fermions
can be constrained in the laboratory experiments on measuring the Casimir force
between two closely spaced test bodies. This force is caused by the zero-point and
thermal fluctuations of the electromagnetic field. It acts between any material
surfaces --- metallic, dielectric or semiconductor \cite{30,31}.
A constraint on the electron-arion coupling constant from old measurements of the
Casimir force \cite{31a} was obtained in \cite{31b}.
The competitive constraints on the coupling constants of axionlike particle to
nucleons from different experiments on measuring the Casimir interaction were
obtained in \cite{32,32a,33,34,35,36,37,38}. All experiments used for obtaining
these constraints have been performed in the separation range below $1~\mu$m.

Starting from 1982 \cite{39}, measurements of the van der Waals and Casimir forces
were also used for constraining the Yukawa-type corrections to Newton's law of
gravitation. These corrections arise due to an exchange of light scalar particles
between atoms of two closely spaced macrobodies \cite{40} and in the
extra-dimensional unification schemes with a low-energy compactification scale
\cite{41,42,43,44}. A review of the most precise measurements of the Casimir
interaction and constraints on non-Newtonian gravity obtained from them can be
found in \cite{20,45}.

In this paper, we obtain new constraints on the coupling constant of axionlike
particles to nucleons and on the Yukawa-type corrections to Newtonian gravity
following from recent experiment on measuring the
differential Casimir force between two
Au-coated bodies spaced at separations from 0.2 to $8~\mu$m \cite{46}.
This experiment was performed by Prof.~R.~S.~Decca by means of a
micromechanical torsional oscillator. The differential Casimir force was
measured between an Au-coated sapphire sphere and the top and bottom of
Au-coated deep Si trenches. The measurement results were compared with the
exact theory using the scattering approach and found to be
in good agreement with it over
the entire measurement range with no fitting parameters under a condition that
the relaxation properties of conduction electrons are not included in
computations. Another theoretical approach, which takes into account the
relaxation properties of conduction electrons, was excluded by the measurement
data over the range of separations from 0.2 to $4.8~\mu$m \cite{46}.

We calculate additional forces arising in the experimental configuration due
to an exchange of two axionlike particles between nucleons of the test bodies,
as well as due to the Yukawa-type correction to Newton's gravitational potential.
Taking into account that no extra force was observed, the constraints on the
masses and coupling constants of axions and on the strength and interaction
range of the Yukawa interaction were found from the extent of agreement between
the measured and calculated Casimir forces.

According to our results, the obtained constraints on axionlike particles are
up to a factor of 4 stronger than those following from other measurements of the
Casimir force. The new constraints on the Yukawa-type interaction are up to a
factor of 24 stronger than those obtained previously from measuring the Casimir force.
Stronger constraints on an axion \cite{35} and non-Newtonian gravity \cite{47}
were obtained only from the experiment by R.~S.~Decca \cite{47} where the Casimir
force was completely nullified.

The paper is organized as follows. In Section 2, we consider the effective
potentials due to an exchange of pseudoscalar and scalar particles. Section 3
is devoted to brief discussion of the recent experiment on measuring
the differential Casimir
force in the micrometer range. In Section 4, we calculate the additional force
due to an exchange of two axionlike particles between nucleons and obtain
constraints on the axion mass and coupling constant. The constraints on the
Yukawa-type correction to Newtonian gravity are found in Section 5.
In Sections 6 and 7, the reader will find the discussion of the obtained results
and our conclusions, respectively.

We use the system of units with $\hbar=c=1$.

\section{Effective Potentials due to Exchange of Pseudoscalar and
Scalar Particles}

\newcommand{\ri}{{\rm i}}

An interaction of the field of originally introduced QCD axions $a(x)$
\cite{9,10,11}, which describes the Nambu-Goldstone bosons, with the
fermionic field $\psi(x)$ is given by the pseudovector Lagrangian
density \cite{12,15}
\begin{equation}
{\cal L}_{\rm pv}(x)=\frac{g}{2m_a}\bar{\psi}(x)\gamma_5\gamma_{\mu}
\psi(x)\partial^{\mu}a(x),
\label{eq1}
\end{equation}
\noindent
where $m_a$ is the mass of an axion, $\gamma_n$ with $n=0,\,1,\,2,\,3,\,4,\,5$
are the Dirac matrices, and $g$ is the dimensionless coupling constant of the
axions to fermionic particles of spin 1/2 (in our case a proton or a neutron).

The axionlike particles introduced in Grand Unified Theories (GUT) interact
with fermions by means of the pseudoscalar Lagrangian
density \cite{12,15,48}
\begin{equation}
{\cal L}_{\rm ps}(x)=-\ri{g}\bar{\psi}(x)\gamma_5\psi(x)a(x).
\label{eq2}
\end{equation}

On a tree level the Lagrangian densities (\ref{eq1}) and (\ref{eq2}) result
in the same action and common effective potential due to an exchange of one
axion or axionlike particle between two nucleons of mass $m$ spaced at a
distance $r=|\mbox{\boldmath$\vec{r}$}_1-\mbox{\boldmath$\vec{r}$}_2|$ \cite{49,50}
\begin{eqnarray}
&&
V_{an}(r;\mbox{\boldmath$\vec\sigma$}_1,\mbox{\boldmath$\vec\sigma$}_2)=
\frac{g^2}{16\pi m^2}\left[
(\mbox{\boldmath$\vec\sigma$}_1\cdot\mbox{\boldmath$\hat{r}$})
(\mbox{\boldmath$\vec\sigma$}_2\cdot\mbox{\boldmath$\hat{r}$})
\left(\frac{m_a^2}{r}+\frac{3m_a}{r^2}+\frac{3}{r^3}\right)\right.
\nonumber \\
&&~~~~~~~~~~~~~~~~
\left.
-(\mbox{\boldmath$\vec\sigma$}_1\cdot\mbox{\boldmath$\vec\sigma$}_2)
\left(\frac{m_a}{r^2}+\frac{1}{r^3}\right)\right],
\label{eq3}
\end{eqnarray}
\noindent
where $\mbox{\boldmath$\hat{r}$}=(\mbox{\boldmath$\vec{r}$}_1-\mbox{\boldmath$\vec{r}$}_2)/r$
is the unit vector directed along the line connecting the two nucleons and
$\mbox{\boldmath$\vec\sigma$}_1,\,\mbox{\boldmath$\vec\sigma$}_2$ are their spins.
Here, we assume equal the coupling constants of an axion to a neutron and
a proton and notate $m$ their mean mass.

Taking into account that the effective potential (\ref{eq3}) is spin-dependent,
the resulting additional force averages to zero when integrated over the
volumes of unpolarized test bodies used in experiments on measuring the
Casimir force. Therefore, the process of one-axion exchange cannot be used
for the axion search in measurements of the Casimir force performed up to
date \cite{31,51} (the proposed measurement of the Casimir force between
two test bodies possessing the polarization of nuclear spins \cite{52}
is, however, quite promising for this purpose).

A process of the two-axion exchange between two nucleons deserves a particular
attention. If the pseudovector Lagrangian density (\ref{eq1}) is used, the
respective effective potential is still unknown. This is because the actual
interaction constant $g/(2m_a)$ is dimensional, and the resulting quantum
field theory is nonrenormalizable (the current status of this problem is
reflected in \cite{53}). However, in the case of the pseudoscalar
Lagrangian density (\ref{eq2}) describing the interaction of axionlike
particles with fermions, the effective potential due to two-axion exchange
is spin-independent and takes the following simple form \cite{40,54,55}:
\begin{equation}
V_{aan}(r)=-\frac{g^4}{32\pi^3m^2}\,\frac{m_a}{r^2} K_1(2m_ar),
\label{eq4}
\end{equation}
\noindent
where $K_1(z)$ is the modified Bessel function of the second kind.

The effective potential (\ref{eq4}) can be summed over all pairs of nucleons
belonging to the test bodies $V_1$ and $V_2$ leading to their total
interaction energy
\begin{equation}
U_{aan}(z)=-\frac{m_ag^4}{32\pi^3m^2}n_1n_2
\int_{V_1}d\,\mbox{\boldmath$\vec{r}$}_1\int_{V_2}d\,\mbox{\boldmath$\vec{r}$}_2
 \frac{K_1(2m_ar)}{r^2},
\label{eq5}
\end{equation}
\noindent
where $n_1$ and $n_2$ are the numbers of nucleons per unit volume
of the first and second test bodies, respectively, and $z$ is the closest
separation distance between them. Finally, from (\ref{eq5}) one arrives to
the additional force acting between the test bodies due to the two-axion
exchange
\begin{equation}
F_{aan}(z)=-\frac{\partial U_{aan}(z)}{\partial z}.
\label{eq6}
\end{equation}

Equations (\ref{eq5}) and (\ref{eq6}) can be used for the search of axionlike
particles in experiments on measuring the Casimir force and for constraining their
parameters.

Similar situation takes place for the Yukawa-type corrections to Newton's
gravitational law which arises due to an exchange of one scalar particle of mass
$m_s$ between two pointlike particles (atoms or nucleons) with masses $m_1$ and
$m_2$ separated by a distance $r$. It is conventional to notate the coupling
constant of the Yukawa interaction as $\alpha G$, where $G$ is the Newtonian
gravitational constant and $\alpha$ is the proper dimensionless Yukawa strength.
Then, the Yukawa-type correction to Newtonian gravity takes the form
\begin{equation}
V_{\rm Yu}(r)=-\frac{Gm_1m_2}{r}\alpha e^{-r/\lambda},
\label{eq7}
\end{equation}
\noindent
where the interaction range $\lambda=1/m_s$ has the meaning of the Compton
wavelength of a scalar particle. As was mentioned in Section 1, the same correction
to Newton's gravitational potential also arises in the extra-dimensional physics
with a low-energy compactification scale  \cite{41,42,43,44}.

The Yukawa-type interaction energy between two macrobodies $V_1$ and $V_2$ arising
due to the potential (\ref{eq7}) is given by
\begin{equation}
U_{\rm Yu}(z)=-\alpha G\rho_1\rho_2
\int_{V_1}d\,\mbox{\boldmath$\vec{r}$}_1\int_{V_2}d\,\mbox{\boldmath$\vec{r}$}_2
\, \frac{e^{-r/\lambda}}{r},
\label{eq8}
\end{equation}
\noindent
where $\rho_1$ and $\rho_2$ are the mass densities
of the first and second test bodies, respectively. In this case,
the additional force acting between two test bodies is equal to
\begin{equation}
F_{\rm Yu}(z)=-\frac{\partial U_{\rm Yu}(z)}{\partial z}.
\label{eq9}
\end{equation}

Both hypothetical forces (\ref{eq6}) and (\ref{eq9}) act on the background of
the Casimir force measured at separations below a few micrometers.
Note that the Newtonian gravitational force
\begin{equation}
F_{\rm gr}(z)= G\rho_1\rho_2\frac{\partial}{\partial z}
\int_{V_1}d\,\mbox{\boldmath$\vec{r}$}_1\int_{V_2}d\,\mbox{\boldmath$\vec{r}$}_2
\, \frac{1}{r}
\label{eq10}
\end{equation}
\noindent
calculated within the same range of separations is much less than the error
in measuring the Casimir force and can be neglected (see below).

\section{Measurements of the Casimir Force in the Micrometer
Separation Range}

All experiments on measuring the Casimir force between two macrobodies used for
obtaining constraints on axionlike particles \cite{32a,33,34,35,36,37,38}
were performed at separations below a micrometer. A single direct measurement
of the Casimir force between two macrobodies at separations up to $8~\mu$m
was reported quite recently  and compared with theory with no fitting parameters
\cite{46}. Below we briefly elucidate the main features of this experiment needed
for obtaining new constraints on the axionlike particles and Yukawa-type corrections
to Newtonian gravity.

In the experiment \cite{46}, the micromechanical torsional oscillator was used to
measure the differential Casimir force between an Au-coated sapphire sphere of
$R=149.7~\mu$m radius and the top and bottom of Au-coated rectangular silicon trenches
of $H=50~\mu$m depth. So large depth of the trenches was chosen in order the
Casimir force between a sphere and a trench bottom (and all the more additional
hypothetical forces) be equal to zero. This means that the actually measured Casimir
force $F_{\rm C}(z)$ acts between a sphere and a trench top. In so  doing, however,
the differential measurement scheme used allowed reaching rather low total
experimental error equal to  $\Delta F_{\rm C}=2.2~$fN at the separation distance
of $z=3~\mu$m. Similar schemes of differential force measurements were previously
used in \cite{47,58,56,57}.

The thicknesses of Au coatings on the sphere and the trench surfaces equal to
$d_{\rm Au}^{(s)}=250~$nm and  $d_{\rm Au}^{(t)}=150~$nm, respectively, were thick
enough in order Au-coated test bodies could be considered as made from solid Au
when calculating the Casimir force.  For technological reasons, before depositing
Au coatings, the sapphire sphere and silicon trench surfaces were also covered
with Cr layers of thickness  $d_{\rm Cr}=10~$nm. These layers do not influence
the Casimir force but should be accounted for in calculations of the additional
forces due to two-axion exchange and the Yukawa-type potential in Sections 4 and
5, respectively.

The Casimir force between an Au-coated sphere and an Au-coated trench top was measured
over the separation region $0.2~\mu\mbox{m}\leqslant z\leqslant 8~\mu$m and compared
with exact theory developed in the sphere-plate geometry using the scattering approach
in the plane-wave representation \cite{59,60,61} and the derivation expansion
\cite{62,63,64,65}.
In doing so the contribution of patch potentials to the measured signal was
characterized by Kelvin probe microscopy. An impact of surface roughness was taken
into account perturbatively \cite{31,51,65a,65b} and found to be negligible.
It was shown \cite{46} that the measurement data are in agreement
with theory to within the total experimental error  $\Delta F_{\rm C}$ over the entire
measurement range if the relaxation properties of conduction electrons are not included in computations. An inclusion of the relaxation of conduction electrons (i.e., using the
dissipative Drude model at low frequencies) results in strong contradiction between
experiment and theory over the range of separations from  0.2 to $4.8~\mu$m.
These results are in line with previous precise experiments on measuring the Casimir
interaction at shorter sphere-plate separations
\cite{57,66,67,68,69,70,71,72,73,74,75,76} and discussions of the so-called Casimir
puzzle \cite{76a,76b,76c} (see also recent approaches to the resolution of this
problem \cite{76d,76e,76f}).

It is necessary to stress that calculation of the Casimir force between the first test
body (sphere) and the flat top of rectangular trenches in \cite{46} was based on first
principles of quantum electrodynamics at nonzero temperature and did not use any
approximate methods, such as the proximity force approximation applied in the most
of previously performed experiments \cite{31,51}, or fitting parameters.
As to the total experimental errors $\Delta F_{\rm C}$, they were found in \cite{46}
at the 95\% confidence level as a combination of both random and systematic errors.

Notice that, according to \cite{ad1,ad2}, one and the same experiment cannot be
used to exclude an alternative model and to constrain the fundamental forces.
This claim was made in 2011 when some of the background effects in the
Casimir force (such as the role of patch potentials) and theoretical
uncertainties (such as an error in the proximity force approximation) were
not completely settled. As was, however, immediately objected in the literature
\cite{ad3,ad4}, such a difference between the excluded and confirmed models of the
Casimir force is of quite another form than a correction to the fundamental force.
This fact allows to constrain the parameters of the latter. After
the seminal experiment by R. S. Decca performed in 2016 \cite{57}, where the
theoretical predictions of two models for the Casimir force differ by up to
a factor of 1000 and one of them was conclusively rejected whereas another one was
confirmed, it became amply
clear that a comparison with the confirmed model can be reliably used for
constraining the fundamental
forces in all subsequent experiments.

When discussing the measure of agreement between experiment and theory, it is necessary
also to take into account the contribution of differential Newtonian gravitational
force between the sphere and the top and bottom of rectangular trenches.
In \cite{46} this force was assumed to be negligibly small. Taking into account that
in Sections 4 and 5 the measure of agreement between experiment and theory is used
for constraining the parameters of axionlike particles and non-Newtonian gravity,
below we estimate the role of Newton's gravitational force more precisely.

The second test body, which is concentrically covered by the rectangular trenches, is
a $D=25.4~$mm diameter Si disc (schematic of the experimental setup is shown
in Figure 1 of \cite{46}).
As mentioned above, both the test
bodies are coated by layers of Cr and Au. In our estimation of the upper bound
for the Newton's gravitational force in this experiment, we assume that both the sphere
and the disc of diameter $D$ and thickness $H$, where $H$ is the trench depth, are
all-gold. By choosing the disc thickness equal to $H$, we disregard the Si substrate
underlying trenches because it does not contribute to the differential gravitational
force between the top and bottom of the trenches (unlike the Casimir and additional
hypothetical forces, the more long-range gravitational force from the trench bottom
cannot be considered equal to zero).

Taking into account that the disc radius $D/2$ is much larger that the sphere
radius $R$, the gravitational force between them is given by \cite{77}
\begin{equation}
F_{\rm gr}=-\frac{8\pi^2}{3}G\rho_{\rm Au}^2HR^3,
\label{eq11}
\end{equation}
\noindent
where $\rho_{\rm Au}=19.28~\mbox{g/cm}^3$ is the density of gold.
Substituting the values of all parameters in (\ref{eq11}), one obtains
$F_{\rm gr}=0.11~$fN. It is seen that the upper bound for the gravitational
force is much less that the experimental error in measuring the Casimir force
which is equal to $\Delta F_{\rm C}=2.2~$fN at $z=3~\mu$m (the variations of
gravitational force depending on whether the sphere is above the top or bottom of
the trenches are well below 0.11~fN). This confirms that the contribution of
Newton's gravitational force in this experiment is very small and can be
neglected.

\section{Constraints on axionlike particles}

The results of experiment \cite{46} on measuring the Casimir force in the
micrometer seperation range allow one to obtain new constraints on the mass of
axionlike particles and their coupling constants to nucleons. As explained in
Section 2, measurements of the Casimir force between unpolarized test bodies can
be used for constraining the parameters of axionlike particles whose interaction
with fermions is described by the Lagrangian density (\ref{eq2}).
For this purpose, one should use the process of two-axion exchange between
nucleons of the laboratory test bodies which leads to the effective potential
(\ref{eq4}) and force (\ref{eq6}). Thus, it is necessary to calculate this force
in the experimental configuration of \cite{46}.

According to Section 3, this configuration reduces to a sapphire (Al$_2$O$_3$)
sphere of radius $R$ (the sapphire density is
$\rho_{{\rm Al}_2{\rm O}_3}=4.1~\mbox{g/cm}^{3}$) coated by the layers of Cr
and Au of thicknesses $d_{\rm Cr}$ and  $d_{\rm Au}^{(s)}$, respectively,
interacting with a Si plate ($\rho_{{\rm Si}}=2.33~\mbox{g/cm}^{3}$) whose
thickness exceeds $H$ and can be considered as infinitely large when
calculating the hypothetical interactions rapidly decreasing with separation.
The plate is also coated by the layer of Cr of thickness $d_{\rm Cr}$ and
by an external layer of Au of thickness  $d_{\rm Au}^{(t)}$ indicated in
Section 3.

The additional force arising in this configuration due to two-axion exchange
can be calculated by (\ref{eq5}) and (\ref{eq6}). The results of this
calculation for homogeneous a sphere and a plate are presented in \cite{38}.
An impact of two metallic layers is easily taken into account in perfect
analogy to \cite{35,37}.
Finally, the additional force due to two-axion exchange in the configuration
of experiment \cite{46} takes the form
\begin{equation}
F_{aan}(z)=-\frac{\pi}{2m_am^2m_{\rm H}^2}\int_1^{\infty}\!\!du
\frac{\sqrt{u^2-1}}{u^2}e^{-2m_auz}X_t(m_au)X_s(m_au),
\label{eq12}
\end{equation}
\noindent
where $m_{\rm H}$ is the mass of atomic hydrogen and the following expressions
for the functions $X_t(x)$ and $X_s(x)$ are obtained:
\begin{eqnarray}
&&
X_t(x)=C_{\rm Au}\Bigl(1-e^{-2xd_{\rm Au}^{(t)}}\Bigr)+
C_{\rm Cr}e^{-2xd_{\rm Au}^{(t)}}\Bigl(1-e^{-2xd_{\rm Cr}}\Bigr)
+C_{\rm Si}e^{-2x(d_{\rm Au}^{(t)}+d_{\rm Cr})},
\nonumber\\
&&
X_s(x)=C_{\rm Au}\Bigl[\Phi(R,x)-e^{-2xd_{\rm Au}^{(s)}}
\Phi(R-d_{\rm Au}^{(s)},x)\Bigr]
\nonumber\\
&&~~~~~~~~~~
+
C_{\rm Cr}e^{-2xd_{\rm Au}^{(s)}}\Bigl[\Phi(R-d_{\rm Au}^{(s)},x)-
e^{-2xd_{\rm Cr}}\Phi(R-d_{\rm Au}^{(s)}-d_{\rm Cr},x)\Bigr]
\nonumber\\
&&~~~~~~~~~~
+C_{{\rm Al}_2{\rm O}_3}e^{-2x(d_{\rm Au}^{(s)}+d_{\rm Cr})}
\Phi(R-d_{\rm Au}^{(s)}-d_{\rm Cr},x),
\label{eq13}
\end{eqnarray}
\noindent
where the function $\Phi$ is defined as
\begin{equation}
\Phi(r,x)=r-\frac{1}{2x}+e^{-4rx}\left(r+\frac{1}{2x}\right).
\label{eq14}
\end{equation}

The numerical coefficients $C$ in (\ref{eq13}) are specific for any material.
They are calculated by the following equation:
\begin{equation}
C=\rho\frac{g^2}{4\pi}\left(\frac{Z}{\mu}+\frac{N}{\mu}\right),
\label{eq14a}
\end{equation}
\noindent
where $Z$ and $N$ are the number of protons and the mean number of
neutrons in an atomic nucleus of a given substance, and  $\mu=M/m_{\rm H}$
with $M$ being the mean atomic or molecular mass.

The values of (\ref{eq14a}) for Au, Cr, Si, and Al$_2$O$_3$, i.e.,
$C_{\rm Au}$, $C_{\rm Cr}$, $C_{\rm Si}$, and $C_{{\rm Al}_2{\rm O}_3}$,
are proportional to $g^2$ with the factors found using the densities of
Au, Si, and Al$_2$O$_3$ indicated above and $\rho_{\rm Cr}=7.15~\mbox{g/cm}^3$.
The values of $Z/\mu$ for Au, Cr, Si, and Al$_2$O$_3$ are equal to
0.40422, 0.46518, 0.50238, and 0.49422, respectively, and of $N/\mu$ to
0.60378, 0.54379, 0.50628, and 0.51412 for the same respective materials
\cite{40}.

Taking into consideration that in the experiment \cite{46} no additional
interaction was observed within the total experimental error
$\Delta F_{\rm C}(z)$, the interaction (\ref{eq12}) should satisfy the
inequality
\begin{equation}
|F_{aan}(z)|\leqslant\Delta F_{\rm C}(z).
\label{eq15}
\end{equation}
\noindent
Using equations (\ref{eq12})--(\ref{eq14}), one can constrain
 from (\ref{eq15}) the possible
values of the axionlike particles mass $m_a$ and their coupling
constant to nucleons $g$.

The numerical analysis of (\ref{eq15}) with account of (\ref{eq12})--(\ref{eq14})
and the values of $\Delta F_{\rm C}$ at different $z$ \cite{46} shows that the
strongest constraints on $m_a$ and $g$ follow at $z=3~\mu$m where
$\Delta F_{\rm C}=2.2~$fN (see Section 4). These constraints are presented by the
line labeled ``new" in Figure~1. For comparison purposes, in the same figure
we show the constraints obtained earlier from the Cavendish-type experiment
\cite{78} (line ``gr$_1$" \cite{79}), from measuring the smallest gravitational
forces  using the torsional oscillator  \cite{80,81,82} (line ``gr$_2$"
\cite{53}), from measuring the effective Casimir pressure \cite{68,69}
(line 1 \cite{32a} ),
from the experiment using a beam of molecular hydrogen \cite{83} (line
``H$_2$" \cite{84}), and from the differential measurement where the Casimir
force was completely nullified \cite{58} (the dashed line \cite{35}).
Note that the figure field above each line is excluded and below each line
is allowed  by the results of respective experiment.
\begin{figure}[!b]
\centering
\vspace*{-8.2cm}
\hspace*{0.cm}\includegraphics[width=20 cm]{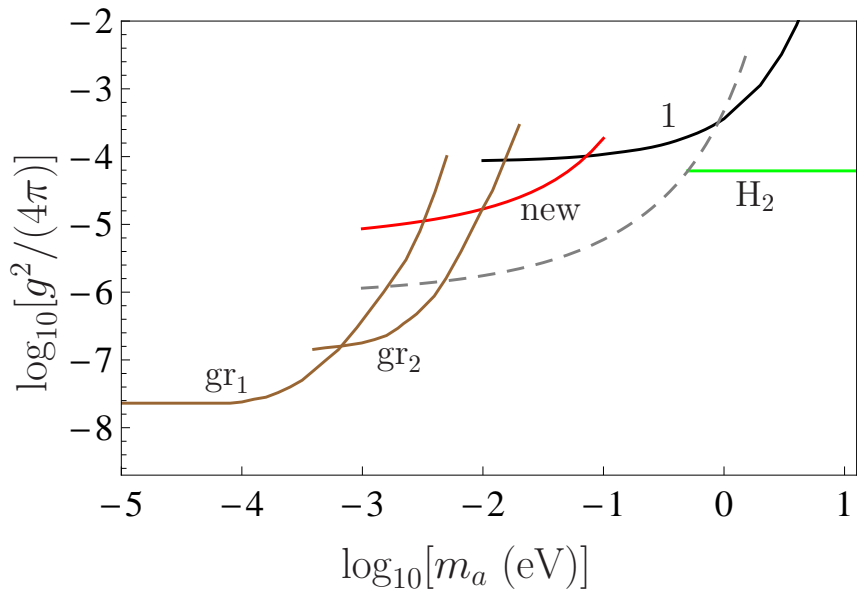}
\vspace*{-11.5cm}
\caption{The strongest constraints on the coupling constant of axionlike
particles to nucleons obtained from different experiments. Lines labeled
``gr$_1$" and ``gr$_2$" are found from the Cavendish-type experiments
and from measuring the minimum gravitational forces, respectively.
Lines labeled ``new" and ``H$_2$" follow from the recent experiment on
measuring the Casimir force in the micrometer separation range and from
the experiment using a beam of molecular hydrogen. Line 1 and the
dashed line are obtained from measuring the effective Casimir pressure and
from the experiment nullifying the Casimir force. The regions above each line
are excluded and below --- are allowed.
\label{fg1}}
\end{figure}

As is seen in Figure~1, in the region of axion masses from 10 to 74~meV
the constraints on $g$ obtained from the recent experiment \cite{46}
(line ``new") are stronger than those following from measuring the effective
Casimir pressure (line 1) and the smallest gravitational forces
(line ``gr$_2$"). The maximum strengthening up to a factor of 4 is reached
for the axion mass $m_a\approx 16~$meV. Thus, in the region of axion masses
indicated above, this experiment leads to stronger constraints than the
previous experiments on measuring the Casimir force \cite{20,45}.
All the more strong constraints shown by the dashed line follow only
from the experiment \cite{58} where the Casimir force was completely
nullified.

\section{Constraints on non-Newtonian gravity}

As was mentioned in Section1, the results of experiment \cite{46} on measuring
the Casimir force in the micrometer separation range can be also used for
constraining the parameters of Yukawa-type corrections to Newtonian gravity.
For this purpose, one should calculate the Yukawa-type force (\ref{eq8}),
(\ref{eq9}) in the experimental configuration using its geometrical
characteristics and densities of all constituting materials presented in
Sections 3 and 4.

The calculation results using (\ref{eq8}) and
(\ref{eq9}) for a homogeneous sphere above a homogeneous plate are
presented in \cite{38}. Here, we modify them in perfect analogy to \cite{35,37}
for taking into account the contributions of Cr and Au layers covering both
a sphere and a trench in this experiment. The result is
\begin{equation}
F_{\rm Yu}(z)=-4\pi^2G\alpha\lambda^3e^{-z/\lambda}Y_t(\lambda)Y_s(\lambda),
\label{eq16}
\end{equation}
\noindent
where
\begin{eqnarray}
&&
Y_t(\lambda)=\rho_{\rm Au}\Bigl(1-e^{-d_{\rm Au}^{(t)}/\lambda}\Bigr)+
\rho_{\rm Cr}e^{-d_{\rm Au}^{(t)}/\lambda}\Bigl(1-e^{-d_{\rm Cr}/\lambda}\Bigr)
+\rho_{\rm Si}e^{-(d_{\rm Au}^{(t)}+d_{\rm Cr})/\lambda},
\nonumber\\
&&
Y_s(\lambda)=\rho_{\rm Au}\Bigl[\Psi(R,\lambda)-e^{-d_{\rm Au}^{(s)}/\lambda}
\Psi(R-d_{\rm Au}^{(s)},\lambda)\Bigr]
\nonumber\\
&&~~~~~~~~~~
+
\rho_{\rm Cr}e^{-d_{\rm Au}^{(s)}/\lambda}\Bigl[\Psi(R-d_{\rm Au}^{(s)},\lambda)-
e^{-d_{\rm Cr}/\lambda}\Psi(R-d_{\rm Au}^{(s)}-d_{\rm Cr},\lambda)\Bigr]
\nonumber\\
&&~~~~~~~~~~
+\rho_{{\rm Al}_2{\rm O}_3}e^{-(d_{\rm Au}^{(s)}+d_{\rm Cr})/\lambda}
\Psi(R-d_{\rm Au}^{(s)}-d_{\rm Cr},\lambda),
\label{eq17}
\end{eqnarray}
\noindent
and the function $\Psi$ is defined as
\begin{equation}
\Psi(r,\lambda)=r-\lambda+(r+\lambda)e^{-2r/\lambda}.
\label{eq18}
\end{equation}

By virtue of the fact that the Yukawa-type force (\ref{eq16}) was not
observed within the limits of the measurement error, it should satisfy
the inequality
\begin{equation}
|F_{\rm Yu}(z)|\leqslant\Delta F_{\rm C}(z).
\label{eq19}
\end{equation}
\noindent
Similar to the case of an additional interaction due to two-axion exchange
between nucleons, the strongest constraints on the parameters of the
Yukawa-type interaction $\alpha$ and $\lambda$ follow from (\ref{eq19})
at $z=3~\mu$m where $\Delta F_{\rm C}=2.2~$fN.

\begin{figure}[!t]
\centering
\vspace*{-8.2cm}
\hspace*{0.cm}\includegraphics[width=20 cm]{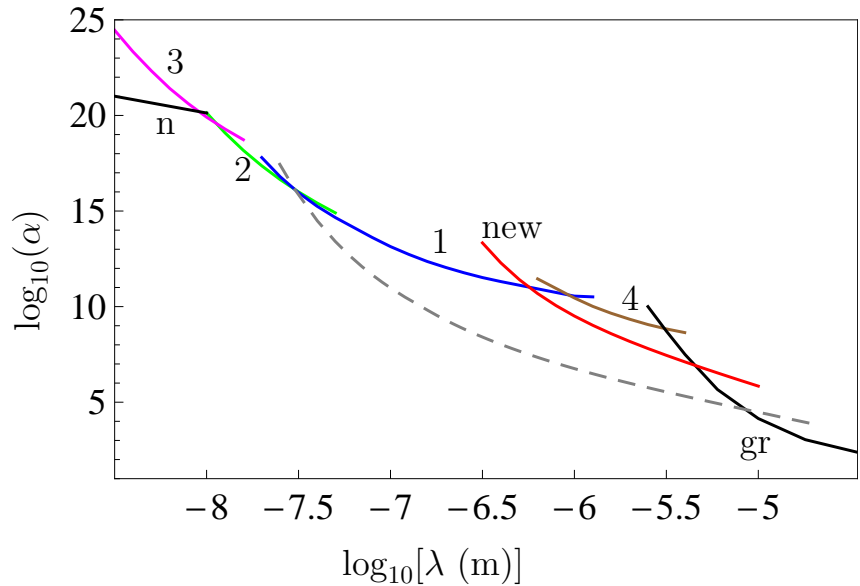}
\vspace*{-11.5cm}
\caption{The strongest constraints on the interaction constant and
interaction range of the Yukawa-type interaction obtained from different
experiments. Lines 1, 2, and 3 are found from measuring the effective Casimir
pressure between two parallel plates, normal, and lateral Casimir forces
between the sinusoidally corrugated surfaces of a sphere and a plate,
respectively. Line labeled ``n" is found from the experiments on neutron
scattering. Lines labeled ``new" and 4 follow from the recent experiment on
measuring the Casimir force in the micrometer separation range and from the
experiment using a torsion pendulum. Line labeled ``gr" and the dashed line
are obtained from the Cavendish-type experiments and from the experiment
nullifying the Casimir force. The regions above each line are excluded
and below each line --- are allowed.
\label{fg2}}
\end{figure}
The obtained constraints are shown by the line labeled ``new" in
Figure~2. For comparison purposes,  in this figure we also show the
constraints following from measuring of the effective Casimir pressure
\cite{68,69} (line 1), of the normal Casimir force between sinusoidally
corrugated surfaces of a sphere and a plate at different orientation
angles of corrugations \cite{85,86} (line 2 \cite{87}), and of the lateral
Casimir force force between sinusoidally
corrugated surfaces of a sphere and a plate \cite{88,89} (line 3 \cite{90}).
Note that somewhat stronger constraints than those shown by the line 3 were
obtained \cite{91} from the experiment on measuring the Casimir force
between crossed cylinders \cite{92} on an undefined confidence level
\cite{31,51}. However, the strongest constraints on the parameters of
Yukawa-type interaction in the region below $\lambda=10^{-8}~$m follow
from the experiments on neutron scattering. They are shown by the line
labeled ``n" \cite{93,94}.
Similar to Figure 1, the regions above each line are excluded
by the results of respective experiment, whereas the regions below each line
are allowed.

We continue description of Figure 2 in the region of larger $\lambda$.
The constraints shown by the line 4 are obtained from measuring the Casimir
force by means of the torsion pendulum \cite{95}. The line labeled ``gr"
indicates constraints on the Yukawa-type interaction following from the
Cavendish-type experiments \cite{78,79,96}. As to the dashed line, which
covers the largest interaction range, it is obtained \cite{58} from the
differential force measurement where the Casimir force was completely
nullified (compare with the dashed line in Figure 1 found from the same
experiment).

As is seen in Figure 2, the constraints labeled ``new" are stronger than
the constraints of lines 1 and ``gr" following from measuring the effective
Casimir pressure and from the Cavendish-type experiments over the
interaction range from 550~nm to $4.4~\mu$m. The maximum strengthening by
up to a factor of 24 is reached at $\lambda=3.1~\mu$m. The obtained
constraints are weaker only as compared to those following from the
experiment where the Casimir force was nullified \cite{58}.

\section{Discussion}

In  this article, we have considered the problems of dark matter axions,
non-Newtonian gravity and constraints on them. As discussed in Section 1,
axions and axionlike particles have gained wide recognition as the most
probable constituents of dark matter. An active search for axions using
their interactions with photons, electrons, and nucleons is under way in
many laboratories all over the world. A major contribution to  the investigation
of interactions between axions and photons in different astrophysical
processes have been made by Prof.~Yu.~N.~Gnedin who suggested several
prospective possibilities for observation of axionlike particles and
constraining their parameters.

Here, we obtain new constraints on the coupling constants of axionlike
particles to nucleons which follow from the recently performed measurement
of the differential Casimir force between Au-coated surfaces of the sphere
and top and bottom of rectangular trenches \cite{46}. The differential
character of this experiment allowed reaching a very high precision and
obtaining the meaningful data up to a very large separation distance of
$8~\mu$m.  The measure of agreement between the obtained data and the
theoretical predictions based on first principles of quantum electrodynamics
at nonzero temperature allowed to find rather strong constraints on the
axionlike particles and non-Newtonian gravity of Yukawa type.

The obtained constraints on the coupling constants of axionlike particles
to nucleons are stronger by up to a factor of 4 than the previously known
ones derived \cite{53} from the gravitational experiments and from measuring
the effective Casimir pressure \cite{33,68,69}. This strengthening holds in
the range of axion masses $m_a$ from 10 to 74~meV.  We have also shown that
the same experiment on measuring the differential Casimir force in the
micrometer separation range \cite{46} results in up to a factor of 24 stronger
constraints on the interaction constant of Yukawa-type interactions as compared
to the ones found previously from measuring the effective Casimir pressure
\cite{68,69},   an experiment using the torsion pendulum \cite{95}, and the
Cavendish-type experiments \cite{78,79,96}. In this case the strengthening holds
in the interaction range from $\lambda=550~$nm to $\lambda=3.1~\mu$m.

Although the obtained constraints are not the strongest ones (the strongest
constraints on both the axionlike particles and the Yukawa-type interaction
in the interaction ranges indicated above were obtained from the experiment
\cite{58} where the Casimir force was nullified), they complement the results
found from previous measurements of the Casimir interaction and can be
considered as their additional confirmation.

The obtained results fall into intensive studies of axionlike particles,
non-Newtonian gravity and constraints on their parameters. In addition to the
literature already discussed in Section 1, it is pertinent to mention a
haloscope search for dark matter axions which excludes some range of the
axion-photon couplings in models of invisible axions \cite{97} and another
haloscope experiment for the search of galactic axions using a
superconducting resonant cavity \cite{98}. The first results of the promising
experiment for searching the dark matter axions with masses below $1~\mu$eV
are reported in \cite{99}. Two more haloscope experiments are performed for
the search of dark matter axions using their interaction with electronic
spins \cite{100} and photons \cite{101}.

A few prospective experiments for constraining the parameters of axionlike
particles and non-Newtonian gravity are suggested in the literature but not
yet performed. Here, one should mention an experiment on measuring the
Casimir pressure between parallel plates at up to $25-30~\mu$m separations
(Cannex) \cite{102,103,103a,104}. An approach for searching dark matter
axions with $m_a<1~\mu$eV using a superconducting radio frequency cavity
is proposed in \cite{105}. Several possibilities for probing the
non-Newtonian gravity in a submillimeter interaction range by means of
temporal lensing \cite{106}, molecular spectroscopy \cite{107}, and neutron
interferometry \cite{107,108} are also discussed. Finally, very recently
the possibility to detect the axion-nucleon  interaction in the Casimir-less
regime by means of levitated  system was proposed \cite{109}. According to
the authors, their approach gives a possibility to strengthen the current
constraints on $g$ by several orders of magnitude in the wide region of
axion masses from $10^{-10}~$eV to 10~eV.

\section{Conclusions}

To conclude, the search for dark matter axions, non-Newtonian gravity and
constraints on their parameters is a multidisciplinary problem of the
elementary particle physics, quantum field theory, gravitational theory,
astrophysics and cosmology. At the moment neither axions nor corrections
to Newton's gravitational law, other than that predicted by the General
Relativity theory, are observed, but more and more stringent constraints
on them are obtained. Keeping in mind that there are very serious reasons
for a creation of axions at the very early stages of the Universe
evolution and for existence of
deviations from the Newton law of gravitation at very
short separations, as predicted by the extended Standard Model,
Supersymmetry, Supergravity and String theory, one may hope that these
predictions will find experimental confirmation in the not too far
distant future.

\funding{
This work was supported by the Peter the Great
Saint Petersburg Polytechnic
University in the framework of the Russian state assignment for basic research
(project N FSEG-2020-0024).
V.M.M.~was also partially funded by the Russian Foundation for Basic Research grant number
19-02-00453 A.
}

\acknowledgments{
V.M.M.~is grateful for partial support by the Russian Government Program of Competitive
Growth of Kazan Federal University. }

\reftitle{References}

\end{document}